\documentclass[aps,prl,twocolumn,superscriptaddress,groupedaddress]{revtex4}
\usepackage{subfig}
\usepackage{graphicx}  
\usepackage{dcolumn}   
\usepackage{bm}        
\usepackage{amssymb}   
\usepackage{slashed}
\usepackage{graphicx}				
\usepackage{amsmath}
\usepackage{mathtools}
\usepackage{tikz,pgf}
\usepackage{comment}
\usepackage{asymptote}
\usetikzlibrary{arrows,backgrounds}
\usetikzlibrary{fit,scopes,calc,matrix,positioning,decorations.pathmorphing}
\usepackage[all]{xy}
\usepackage{yfonts}

\newcommand{\Bracket}[1]{\ensuremath{\left\langle#1\right\rangle}}

\begin{document}
\title{Spacetime vacuum as a correlated quantum channel, dual to gravitational memory}
\author{Andrei T. Patrascu}
\address{ELI-NP, Horia Hulubei National Institute for R\&D in Physics and Nuclear Engineering, 30 Reactorului St, Bucharest-Magurele, 077125, Romania\\
email: andrei.patrascu.11@alumni.ucl.ac.uk}
\begin{abstract}
In this note I will argue that the spacetime vacuum of any gauge theory plays the role of a correlated quantum channel and that the concept of a correlated quantum channel is dual to gravitational memory. The existence of memory in the case of other gauge theories is discussed and a similar duality is identified suggesting that the vacuum of any theory could play such a role. This can play a role in the resolution of the black hole information paradox.
\end{abstract}
\maketitle
The black hole information paradox arises from the fact that the region where the particles are being created around the black hole horizon is always far away in a causal sense from the in-falling matter that the particles should in principle encode. Therefore, the escaping quanta are presenting only a perturbatively small and totally insufficient entanglement with the in-falling matter. This gives then rise to difficulties in treating black holes unitarily [1], [2], [3]. While we do assume that entanglement and hence quantum information must play a role in the resolution of this problem, the exact mechanism remains unknown [4]. 
There are however two phenomena that are at play in any vacuum in spacetime that are relevant to this problem. First, the existence of correlated quantum channels [5], [6], [7], which I propose to be at the very foundation of spacetime. Second, a proposed duality between gravitational memory and correlated quantum channels. In fact such a postulated duality is expected to exist for the memory associated to any non-linear gauge field theories and the correlated quantum channels associated to their respective vacua. The gravitational case however is more notorious due to its long range effects that remain measurable, as opposed to the more difficult limits of Yang Mills or non-linear QED. 
To make matters clearer, let us consider what a correlated quantum channel is and how it can be interpreted. 
\par Historically, it has been noted that entanglement between the endpoints of a communication channel can increase its classical capacity [8]. As classical information can be encoded in terms of the associated entropy, quantum information has been associated in such a case with entanglement entropy, and in principle it only reflects the information that is not recoverable by only one side of the system, no matter what local measurements one performs on it. Clearly, if the information we wish to transmit is partly spread between the two sides in the form of an entanglement construction, less information has to be transmitted through the channel to recover the transmitted message and hence the capacity of the channel will increase. This is the basic reason why quantum communication is considered to be at least more efficient than communication through purely classical channels.
\par In all these situations the channel has been regarded as mainly passive, with structureless noise at best. From classical information theory we know that information is the equivalent of randomness. This observation is fascinating not only because it intuitively associates entropy to information, but because it defines transmitted information as something that cannot be simply inferred from the measurements without something that is essentially unpredictable on the side of the receiver. That this definition implies that information should be fundamentally quantum I discussed previously. An additional important element however is the channel which in many circumstances cannot be seen as a passive contributor to the communication process. Indeed, thinking of the information channel as a correlation sensitive instrument, we may assign to it a certain type of correlation between past transmitted data and present or future transmitted data. This type of correlation could be associated with our intuitive concept of memory, and even stronger, if we consider just transmitting signals through spacetime, with gravitational-type memory. Indeed, we can entangle information sent at some instant through the channel with information sent at a later instant, and with even later information, and so on, creating a chain of entangled uses of the channel. In that sense, the channel will not at all be neutral to the information transmitted through it, but instead will contribute and "get used" to the types of information that appear to be the likeliest. We obtain a quantum memory-enhanced channel for which it has been shown that the amount of transmitted information can be at most 
\begin{equation}
C_{n}=\frac{1}{n}sup_{\epsilon}I_{n}(\epsilon)
\end{equation}
where $\epsilon=\{P_{i}, \pi_{i}\}$ with $P_{i}\geq 0$, $\sum_{i}P_{i}=1$ is the input ensemble of states $\pi_{i}$, transmitted with a priori probabilities $P_{i}$, of $n$ qubits, while $I_{n}(\epsilon)$ is the mutual information and $n$ is the number of uses of the channel. 
This is a simple example of a channel that can itself "learn" by entangling past and present data that passes through it. Of course this type of channel can be expanded to an entangling network, which could be assumed to be spacetime itself, but I will make this observation in a more intuitive sense differently. Up to now we are discussing about some arbitrary quantum channel, which we could construct somehow. However, the spacetime vacuum itself can be seen as a network of such channels. Both entanglement harvesting processes and the Cutkosky's theorem are suggestive to this. Let us first start with Cutkosky's theorem. Let our theory be
\begin{equation}
S=\int d^{D}x (\frac{1}{2}(\partial\phi)^{2}+m^{2}\phi^{2})-\frac{g}{3!}\phi^{3})
\end{equation}
and after summing up the 1PI insertions we obtain 
\begin{equation}
i \mathcal{D}_{\phi}(q)=\frac{i}{q^{2}-m^{2}-\Sigma(q)+i\epsilon}
\end{equation}
with $\Sigma(q)$ the two point vertex and $i\epsilon$ the small imaginary deformation. 
If we want to calculate the two point vertex, we obtain a diagrammatic expansion with the leading contribution coming from the one loop diagram 
\begin{equation}
i\Sigma_{1 loop}(q^{2})=(ig)^{2}\int d^{D}k\frac{i}{k^{2}-m^{2}+i\epsilon}\frac{i}{(q-k)^{2}-m^{2}+i\epsilon}
\end{equation}
smashing the inner lines on shell, we get $q$ real and actually realised states for the scalar field, namely timelike $q^{\mu}$ and $q_{0}>m$. The real part of $\Sigma$ will produce a shift in the mass. The imaginary part of this function will represent the decay rate $\Gamma=\frac{Im(\Sigma)}{m}$. The energy of the particle is shifted away from the real axis in the well known and simple way:
\begin{equation}
\sqrt{m^{2}+i Im\Sigma(m^{2})} \sim m+i\frac{Im\Sigma(m^{2})}{2m}
\end{equation}
If we Fourier transform to the real time we get an amplitude for propagation in time of a state with complex energy, which obviously is defined like a state
\begin{equation}
\psi(t)\sim e^{-i \epsilon t}
\end{equation}
and the norm of this state would be
\begin{equation}
||\psi(t)||^{2}\sim ||e^{-i(E-i \frac{1}{2}\Gamma)t}||^{2}=e^{-\Gamma t}
\end{equation}
with $\Gamma \sim Im(\Sigma(m^{2}))/m$ being the rate of decay of the norm of the single particle state. The Fourier transform of the wavefunction $\psi(t)$ simply produces our decay resonance
\begin{widetext}
\begin{equation}
F(\omega)=\int dt e^{-i\omega t} \psi(t) = \int_{0}^{\infty} dt e^{-i \omega t}e^{i(M-\frac{1}{2}\Gamma)t}=\frac{1}{i(\omega-M)-\frac{1}{2}\Gamma}
\end{equation}
\end{widetext}
We are not surprised. This is the decay resonance, which is a Lorentzian and has width $\Gamma$. 
In the hope I am not filling up this article with trivia, I think it is still important to review how we get to the Cutkosky cutting rules, and we are not there yet. 
At the one loop level we employ the principal part of an integral $\mathcal{P}$ to re-write 
\begin{equation}
\frac{1}{k^{2}-m^{2}+i\epsilon}=\mathcal{P}\frac{1}{k^{2}-m^{2}}-i\pi\delta(k^{2}-m^{2})=\mathcal{P}-i\Delta
\end{equation}
and we get
\begin{equation}
Im(\Sigma_{1 loop}(q))=-g^{2}\int d\phi(\mathcal{P}_{1}\mathcal{P}_{2}-\Delta_{1}\Delta_{2})
\end{equation}
where 
$d\phi=dk_{1}dk_{2}(2\pi)^{D}\delta(k_{1}+k_{2}-q)$
After some simple algebra, we can represent the 1-loop self energy in terms of real space propagators, and we obtain the amplitude for the creation of two $\phi$ excitations from a single field, their propagation to a point $x$ and then their annihilation into a single field $\phi$ again. 
Restricting to only the positive energy components satisfying the momentum conservation conditions we get
\begin{widetext}
\begin{equation}
Im(\Sigma)=g^{2}\int d\phi \theta(k_{1}^{0})\theta(k_{2}^{0})\Delta_{1}\Delta_{2} = \frac{g^{2}}{2}\int \frac{d^{D-1}k_{1}}{2\omega_{k_{1}}}\frac{d^{D-1}k_{2}}{2\omega_{k_{2}}}(2\pi)^{D}\delta^{D}(k_{1}+k_{2}-q)
\end{equation}
\end{widetext}
This is the final form of the Cutkosky cutting rules and it gives the imaginary part of a Feynman diagram. They dictate the decay process as well as the decay amplitude and the width of the decay resonance. Basically, given a Feynman diagram with an inner loop (or several inner loops, but 1PI) that is properly amputated, we can draw a line through the diagram separating initial and final states, cutting a series of internal propagators. We replace all cut propagators by 
\begin{widetext}
\begin{equation}
\theta(p^{0})\pi\delta(p^{2}-m^{2})=\theta(p^{0})\frac{\pi\delta(p_{0}-\epsilon_{p})}{2\epsilon_{p}}
\end{equation}
\end{widetext}
If we smash the intermediate particles on-shell, the amplitude becomes imaginary, which means simply that the decay process of the initial particle into the virtual particles turning real is governed by the imaginary part of the amplitude. What is interesting to notice is that those particles in fact are entangled. This results quite obviously from the observation that the vacuum state is (in a sense by definition) entangled, and that vacuum contains all the loops that we can cut through and obtain decay processes which result in entangled particles. So, either we think in terms of vacuum as an entangled state from which we extract entanglement, or we think of it as a contribution from loop diagrams that, when placed on-shell provide entangled decay products. Simply this statement probably wouldn't have warranted leading the reader through the whole Cutkoski process, if it wasn't for the observation that the process described in the decay is a direct result of unitarity on one side, and that there is no cut-off dependence of the imaginary part, as that would be at odds with unitarity as well. 

This completes the argument that the vacuum can be seen as a set of entangled states, and that it can play the role of an entangled substructure in which each process is entangled not only within its own loop, but also with the next carrier passing through our channel. Actually, up to now we only can see that each channel is entangled within itself i.e. the vacuum carrier production leaves us with more entanglement than what we started with, but not yet with the next carrier. To see that, we have to understand that there are categories of diagrams that have non-unique ways in which they can be cut. For example a box diagram, representing for example photon-photon scattering with the exchange of electrons and positrons, (a situation of a channel carrier interacting with the background of the channel) there are more than one way to cut the diagram. Those result in discontinuities in various kinematical variables and the imaginary part has to be obtained by adding those together. In that case however there will be some exchanged particles with spacelike momentum which cannot in principle be cut but will play a role in adding up the diagram cuts. In these channels we can see entanglement between various instances of the diagram, or, thinking in a causal structure, the entanglement of one quantum channel carrier with the next.
So, if we look at the vacuum of a theory, we will have to consider it as a correlated (or correlation sensitive) quantum channel, and that finally amounts to a channel with some form of memory. The memory in a quantum channel appears like some form of correlation between events on distinct points in a time-series. But there is one interesting property of non-linear theories including Gravity, harder to observe in Yang Mills theories, or (non-linear) electromagnetism, where such a memory does not quite appear in a manifest way. This is called the "memory effect". In gravitational physics, this effect amounts in the change of various properties of a system of detectors traversed and left by a gravitational (shock)wave. The transition of such a wave leaves a persistent effect on spacetime in the form of a "memory" and equivalently in the form of a highly correlated quantum channel (or correlation sensitive quantum channel). Basically a gravitational wave passing through spacetime will leave a long-lived deformation of the spacetime that is the result of the non-linearity of the Einstein's field equations. A similar non-linearity can be found in the situation of analysing the effective photon-photon scattering via the square diagram, where interestingly enough, the cutting procedures for finding the decay process take a non-trivial form in which entanglement appears in a channel that shows spacelike propagation. The same process can be calculated in Yang Mills theories, although it has the well known obstructions due to the phase transitions of Yang Mills, and in the case of the electroweak theory we have the difficulty arising from the Higgs mechanism coming in our way. However, the effect is rather universal. 
The Memory effect is not purely non-linear. A linear counterpart appearing in linearised gravitational wave descriptions was well known as the result of the global change of the second time derivative of the quadrupole moment of the source i.e. in a change of the linear momenta of the constituent bodies. However, ref. [9] has shown that any gravitational wave burst has gravitational memory, due to non-linear effects. Given the various polarisations of the gravitational waves, this effect leads to the fact that for some of the polarisation components we will have a non-zero value for 
\begin{equation}
\Delta h_{+ x}=\lim_{t\rightarrow \infty} h_{+ x}(t) - \lim_{t\rightarrow -\infty}h_{+ x}(t)
\end{equation}
where the observer will make its measurements at time $t$. That means a gravitational wave will not bring the detector back to its original state. 
Such a memory effect is dual in a sense, with the effect of a correlated quantum channel, and such a correlated quantum channel can be constructed from vacuum diagrams with internal loops, unmeasurable in flat space but relevant in curved space, and generally relevant in cosmological contexts as well as in the context of black holes. Again "non-measurability" is a bit of a misnomer, as the entanglement structure of the vacuum can in principle be detected if one manages to create particles from the vacuum and measure their entanglement, which leads to a relatively certain probe of the idea of a correlated spacetime, as implied here. 
A simple approach to gravitational memory relies on the development of gravitational wave polarisations in a sum over various (l,m) modes
\begin{equation}
h_{+}-ih_{x}=\sum_{l=2}^{\infty}\sum_{m=-l}^{l}h^{lm}\cdot _{-2}Y^{lm}(\Theta, \Phi)
\end{equation}
We have done a spin-weighted spherical harmonic expansion. In a multipolar expansion of the gravitational field, the modes can be written as a complex expression combining the radiative mass and the current multipoles
\begin{equation}
h^{lm}=\frac{1}{\sqrt{2}R}[U^{lm}(T_{R})-iV^{lm}(T_{R})]
\end{equation}
where $R$ is the distance from the source to the observer, $T_{R}$ is the retarded time, and the respective moments $U^{lm}$ and $V^{lm}$ are constructed from the symmetric trace free tensors of rank $l$
\begin{equation}
\begin{array}{cc}
U^{lm}=A_{l}\mathcal{U}_{L}\mathcal{Y}_{L}^{lm*}, & V^{lm}=B_{l}\mathcal{V}_{L}\mathcal{Y}_{L}^{lm*}
\end{array}
\end{equation}
with $A_{l}$ and $B_{l}$ parameters depending on $l$ and $\mathcal{Y}_{L}^{lm}$ spherical harmonics. 
The radiative momenta $U$ can be defined in terms of source multipole moments which are defined as integrals over the stress energy pseudotensor describing matter and gravitational fields of the source. Making quick use of references [9], [10], this is
\begin{widetext}
\begin{equation}
U_{lm}=I_{lm}^{l}+2\mathcal{M} \int_{-\infty}^{T_{R}}[ln(\frac{T_{R}-\tau}{2\tau_{0}})+\kappa_{l}]+I_{lm}^{l+2}(\tau)d\tau + U_{lm}^{(non-lin)}+\mathcal{O}(2.5 PN)
\end{equation}
\end{widetext}
where $I^{(l)}_{lm}$ is the $l-th$ time derivative of the mass source moment $I_{lm}$, the integral term being an order $1.5 PN$ tail term, $\mathcal{M}$ is a mass monopole moment, the rest of the quantities are known and the $U^{(non-lin)}$ term is the non-linear memory effect. 
Without going into much detail as the precise calculation of the memory effect for this or that phenomenon is not truly relevant, but following the calculations of [10] for a waveform of a hyperbolic binary, we get the leading order multipolar contribution to the polarisations as 
\begin{equation}
h_{+}-ih_{x}\sim \sum_{m=-2}^{m=2}\frac{I_{2m}^{(2)}}{R\sqrt{2}} \cdot _{-2}Y^{lm}(\Theta, \Phi)
\end{equation}

or we can obtain the linear memory for an unbound system by solving the linearised harmonic gauge Einstein field equations for the space-space piece of the metric perturbation $h_{jk}$
\begin{equation}
\Box\bar{h}_{jk}=-16 \pi T_{jk}
\end{equation}
where $T_{jk}$ is the stress energy tensor of $N$ gravitationally unbound particles of masses $M_{A}$ and constant velocities $v_{A}$, and $\bar{h}_{jk}$ is the trace reversed metric perturbation, forming a flat space wave equation by means of the d'Alembertian. Solving this [10] leads to 
\begin{equation}
\Delta h_{jk}^{TT}=\Delta\sum_{A=1}^{N}\frac{4M_{A}}{R\sqrt{1-v_{A}^{2}}}[\frac{v_{A}^{j}v_{A}^{k}}{1-v_{A}\cdot N}]^{TT}
\end{equation}
$N$ pointing towards the observer and $\Delta$ represents the difference between late and early times.

Next, the non-linear memory effect, comes from a contribution to the radiative mass multipole moments $U_{lm}$ generated by the energy flux of the radiated gravitational waves. 
The correction term to the gravitational wave [10] is 
\begin{equation}
\delta h_{jk}^{TT}=\frac{4}{R}\int_{-\infty}^{T_{R}}dt'[\int \frac{dE^{gw}}{dt'd\Omega'}\frac{n'_{j}n'_{k}}{(1-n'\cdot N)}d\Omega']^{TT}
\end{equation}
using the same notations. We notice that we are integrating over time hence we generate a contribution that depends on the entire past history of the source. If the unbound objects in the system are considered to be soft gravitons then we can relate the non-linear memory to the linear one. 
This time integral is important as it generates a correlation between different points in a time-series, basically transforming our contribution to the gravitational field correction into a highly time-correlated quantum channel, in which memory is given by means of gravitational memory. 
This suggests a duality between the two effects: on one side gravitational memory appears as a construction that permanently alters spacetime in the sense of leaving a long-term imprint of the passage of the gravitational wave, on the other side showing that the spacetime itself is a correlation sensitive information channel. Whether that channel is quantum is probably the most intriguing part of this duality. In fact the quantum channel could be constructed from the loop diagrams described in the first part of this paper, in which the loops are highly entangled, including between different time-series points. 
Such channels with consecutive noise are encoded by qubits having a unitary evolution in time. Of course, that would amount to remembering the evolution of the S-matrix in the sense of 
\begin{equation}
S_{fi}=\Bracket{f,e^{-iH\cdot T},i}=(1+i\mathcal{T})_{fi}
\end{equation}
and due to unitarity we would have 
\begin{equation}
H=H^{\dagger} \rightarrow 1=S\cdot S^{\dagger}
\end{equation}
and hence the optical theorem
\begin{equation}
2 Im\mathcal{T}=i(\mathcal{T}^{\dagger}-\mathcal{T})=\mathcal{T}^{\dagger}\mathcal{T}
\end{equation}
and we can re-write it as 
\begin{equation}
2 Im \mathcal{T}_{fi}=\sum_{n}\mathcal{T}^{\dagger}_{fn}\mathcal{T}_{ni}
\end{equation}
and at the one loop level the intermediate states will be two particle states. This means that at a microscopic level integrating over the entire time interval of the non-linear gravitational memory region, we will implicitly have to integrate on each time-slice of that memory over all affected loop diagrams hence generating (or, better, manifestly taking into account) the inner loop entanglement. Then we will move at the next time slice and do the same until all slices are accounted for and we would obtain the integral contribution to the non-linear gravitational memory effect. This sounds somehow artificial because we define the loop diagrams over the spacetime of the gravitational wave, but a simpler approach is to think at non-linear memory effects in quantum electrodynamics. There we also have a Christodoulou type memory effect where the quantity of importance is the change in velocity of the test particle considering an electric field. The effect is persistent at early and late times after integration over time. This effect can be calculated classically, but the quantum field theoretical calculation is directly related to the outcome by means of non-linear QED effects emerging from square diagrams or other loop diagrams. Such calculations have been performed, including the calculation of an effective QED Lagrangian in a constant external field $F_{\mu\nu}$. At one loop for example in spinor QED the Euler Heisenberg Lagrangian is 
\begin{widetext}
\begin{equation}
\mathcal{L}_{spin}^{(1)}(F)=-\frac{1}{8\pi^{2}}\int_{0}^{\infty}\frac{dT}{T^{3}}e^{-m^{2}T}[\frac{(eaT)(ebT)}{tanh(eaT) tan(ebT)}-\frac{1}{3}(a^{2}-b^{2})T^{2}-1]
\end{equation}
\end{widetext}

where $T$ is the proper time of the loop fermion, $m$ is its mass and $a$, $b$ are Maxwell invariants related to $E$ and $B$. This is of course an approximation in which all photon energies can be neglected compared to the electron mass, but it contains the one-loop information about the photon S-matrix. In order to obtain a good estimation of the entanglement, further integration must be performed at higher loops, but it is enough to notice that entanglement will emerge at the level of the inner loops, or if measuring a scattering experiment of this type, in which the inner loops will entangle the emerging photons. The integration over the electric field in the memory time integral will become a multiple series of loop integrals at the quantum level, leading to a highly entangled, hence highly correlated and correlation sensitive quantum channel. That is where memory comes from. 
Now is probably the time to discuss the black hole information problem using plain words.
The phenomenon that takes place around the horizon of a black hole is to have a spacelike slice that extends across the horizon, where inside the horizon the timelike directions and spacelike directions become interchanged. This leads to the connective region around the horizon to be stretched enhancing Fourier modes from around the Planck scale (or from high energies) resulting in particle production. The main problem is that for an object that falls inside the black hole to have its information restored by means of radiated hawking particles from this region, it should at least be in causal contact with the Hawking quanta at some point for entanglement to occur. No matter how the problem is analysed, there is no mechanism in which enough entanglement between the in-falling object and the emerging quanta in the stretching region can be formed. The entanglement is either zero or vanishingly small. Come in correlation sensitive channels. If spacetime itself is a correlation sensitive channel as proposed above and as showed by means of the gravitational (and other non-linear) memory effects and their connection to entanglement of the vacuum, then it is sufficient to have the gravitational (or other gauge) field imprint left by the in-falling object on the vacuum seen as an entangled (correlation sensitive channel) to obtain information about the in-falling object. 
The correlated quantum channel has the possibility of exchanging the entanglement between the produced Hawking quanta with the entanglement between the Hawking quanta and the residue that left an imprint on the quantum correlated channel, which in this case is the stretched spacetime path. Therefore, as more matter falls into the black hole, more information can be extracted from the in-falling matter by means of Hawking radiation. 
\par Whether we want to call it gravitational memory, or a correlated channel, it is the channel itself (or otherwise spacetime itself) that will reproduce the information. That kind of encoding is obviously global in nature, and hence it is not a local effect on the Hawking radiation, hence entanglement entropy and a larger part of the radiation has to be considered. Therefore this hypothesis is consistent with the rest of the analysis suggesting that no local or small perturbation can explain the black hole information paradox. 
\section{notes}
On behalf of all authors, the corresponding author states that there is no conflict of interest.\\
My manuscript has no associated data.

\end{document}